\documentclass[showpacs]{article}
\usepackage{graphicx}

\begin{document}

\title{\textbf{Cooper pairs as bosons}}
\author{M. de LLANO$\dagger\S$, F.J. SEVILLA$\S$ and S. TAPIA$\ddagger$ \\
\\
$\dagger $Texas Center for Superconductivity, University of Houston \\
Houston, TX 77204 USA \&\\
Instituto de Investigaciones en Materiales, UNAM\\
04510 M\'{e}xico DF, Mexico\\
$\S $Consortium of the Americas for Interdisiplinary Science\\
University of New Mexico\\
Albuquerque, NM 87131, USA\\
\\
$\ddagger$Instituto de F\'{\i}sica, UNAM, 01000 M\'{e}xico, DF, Mexico}
\maketitle

\begin{abstract}
Although BCS pairs of fermions are known not to obey Bose-Einstein (BE)
commutation relations nor BE statistics, we show how Cooper pairs (CPs),
whether the simple original ones or the CPs recently generalized in a
many-body Bethe-Salpeter approach, being clearly distinct from BCS pairs at
least obey BE statistics. Hence, contrary to widespread popular belief, CPs
can undergo BE condensation to account for superconductivity if charged, as
well as for neutral-atom fermion superfluidity where CPs, but uncharged, are
also expected to form.
\end{abstract}

\hspace{0.86cm}{\small \textit{Keywords}: Cooper pairs; Bose-Einstein
condensation; superconductors; superfluids} 

\section{Introduction}

A recent electronic analog of the Hanbury Brown-Twiss photon-effect
experiment Samuelsson and B\"{u}ttiker\ \cite{Samuelsson}\ strongly suggest
electron pairs in a normal/superconductor junction to be bosons, although
further work seems needed to provide compelling empirical proof. On the
other hand, assuming Cooper pairs (CPs) to be bosons, it was recently proved %
\cite{Tolma,PLA2} within a generalized Bose-Einstein condensation (BEC)
model that includes the BCS-Bose crossover theory \cite{Friedel}-\cite{PRB95}
as a special case, that a BCS condensate is precisely a Bose-Einstein (BE)
condensate consisting of equal numbers of particle- and hole-CPs, at least
for the Cooper/BCS \cite{Cooper,BCS}\ model interfermion interaction in the
limit of weak coupling when the crossover picture reduces to BCS theory.
This is significant since BCS argued (Ref. \cite{BCS}, footnote 18) that ``%
\textit{our transition is not analogous to a Bose-Einstein condensation}
(BEC).'' Somewhat later, Bardeen wrote \cite{PT} that ``\textit{...the
picture by Schafroth (1955)...of electron pairs...which at low temperature
undergo a BEC, is not valid.}'' In contrast, both Feynman \cite{Feymann}\
and Josephson \cite{Josephson}\ referred without proof to CPs\ as bosons.

On the other hand, a long-standing common objection to attempts \cite%
{Tolma,PLA2,CMT02,deLlRev} to unify BCS and BEC theories for a description
of \textit{superconductivity} has been that CPs, strictly speaking, cannot
be considered bosons. In this Letter we establish a clear distinction
between BCS pairs and CPs, showing that while the former are not bosons the
latter very definitely are, at least insofar as they obey BE statistics. The
distinction, though elementary, is hardly if ever mentioned in the
superconductivity literature, though it appears not to be as unacceptable
among the neutral-fermion \textit{superfluidity }community where it is also
applied, albeit with a different interfermion interaction, to the BCS and
Cooper pairs as in liquid $^{3}$He \cite{He3} and in ultracold trapped
alkali Fermi gases such as $^{40}$K \cite{Holland} and $^{6}$Li \cite{Li6}.
Recent studies \cite{Holland01}-\cite{Stringari02} have also appeared\
addressing superfluidity in degenerate Fermi gases. Indeed, a BE-like
condensate of dimers formed from neutral $^{40}$K\ \cite{Jin}\ or $^{6}$Li %
\cite{Hulet}-\cite{Ketterle} fermionic atoms via magnetically-tunable
``Feshbach resonances'' has only recently finally been observed.

\section{Cooper pairs}

The original or ordinary CPs \cite{Cooper} emerge by solving a Schr\"{o}%
dinger-like equation in momentum space for two \textit{particles }above the
ideal-Fermi-gas (IFG) sea of the remaining $N-2$ system fermions. If
two-hole pairs are also included, the extended Cooper eigenvalue equation
yields \cite{bts}\cite{AGD}\ purely imaginary values. This precludes
defining CPs relative to an IFG sea. However, a more both general and
self-consistent treatment starts with an unperturbed Hamiltonian
representing not the Fermi sea but BCS-correlated ground state in the
Bethe-Salpeter (BS)\ integral equations \textit{with }hole propagation. When
solved \cite{Honolulu}-\cite{1D}\ in the ladder approximation for bound
two-particle and two-hole states, non-purely-imaginary eigenvalues are
restored for what has been called the ``moving CP'' solution of the BS
equations. The BS results in 3D \cite{Honolulu}, 2D \cite{EPJD}, and 1D \cite%
{1D}\ are consistent with each other, as expected.

Unlike the Fermi-atom case, the interfermionic interactions in (especially
cuprate) superconductors are a deep, problematic and highly controversial
issue \cite{Lanzara}-\cite{Kulic}. So, we confine ourselves to an electron
gas interacting pairwise via the ingeniously simple Cooper/BCS model
interfermion interaction \cite{Cooper,BCS} 
\begin{equation}
V_{\mathbf{kk^{\prime }}}=\left\{ 
\begin{array}{cl}
-V\, & \hbox{if }k_{F}\,\ \ {{{\rm or}\thinspace }\ \ }\mathrm{\sqrt{%
k_{F}^{2}-k_{D}^{2}}<|\mathbf{k}\pm \frac{1}{2}\mathbf{K}|,\,|\mathbf{k}%
^{\prime }\pm \frac{1}{2}\mathbf{K}|<\sqrt{k_{F}^{2}+k_{D}^{2}}} \\ 
0 & \hbox{otherwise,}%
\end{array}%
\right.   \label{BCSint}
\end{equation}%
where $V_{\mathbf{kk^{\prime }}}\equiv L^{-d}\int d\mathbf{r}\int d\mathbf{r}%
^{\prime }e^{-i\mathbf{k}\cdot \mathbf{r}}V(\mathbf{r},\mathbf{r}^{\prime
})e^{i\mathbf{k}^{\prime }\cdot \mathbf{r}^{\prime }}$ is the double Fourier
transform of $V(\mathbf{r},\mathbf{r}^{\prime }),$ the (possibly nonlocal)
interaction in $d$-dimensional coordinate space, with $\mathbf{r}$ the
relative coordinate of the two electrons. Here $V>0$, and $\hbar \omega
_{D}\equiv \hbar ^{2}k_{D}^{2}/2m$ is the maximum energy of a phonon
associated with the vibrating ionic lattice underlying the electron gas,
while $E_{F}\equiv \hbar ^{2}k_{F}^{2}/2m$ is the Fermi energy and $m$ is
the effective electron mass. In Ref. \cite{Cooper} there are no hole pairs
(viz., below $E_{F}$)\ so the lower limit in (\ref{BCSint}) was $k_{F}$,
while in Ref. \cite{BCS} it is $\sqrt{k_{F}^{2}-k_{D}^{2}}$. In the simpler
Cooper case it means that two electrons with momentum wavevectors $\mathbf{k}%
_{1}$ and $\mathbf{k}_{2}$\ interact with a constant net attraction $-V$
when the tip of 
\begin{figure}[t]
\begin{center}
\includegraphics{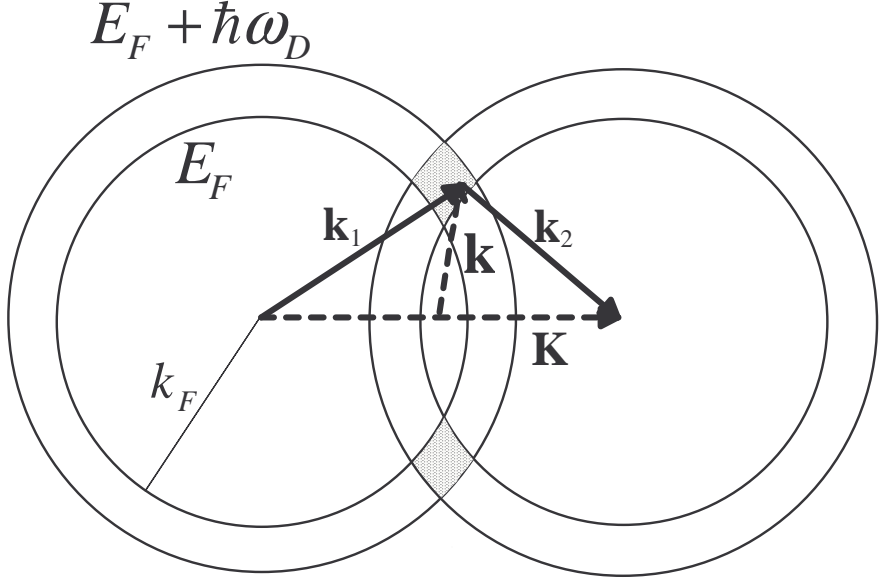}
\end{center}
\caption{Cross-section of overlap volume in $k$-space (shading) where the
tip of the relative pair wavevector $\mathbf{k}$ $\mathbf{\equiv }$ ${\frac{1%
}{2}}\mathbf{({k}_{1}-{k}_{2})}$ must point for the attractive Cooper/BCS
model interaction (\ref{BCSint}) to be nonzero for a pair of total (or
center-of-mass) momentum $\hbar \mathbf{K}$ ${\equiv \hbar \mathbf{(k}}_{1}+%
\mathbf{k}_{2})$. The tail of $\mathbf{k}$ is at the midpoint of $\mathbf{K}$%
.}
\label{BCSandCPsf1}
\end{figure}
their relative-momentum wavevector $\mathbf{k\equiv {\frac{1}{2}}(k}_{1}-%
\mathbf{k}_{2}\mathbf{)}$ points anywhere inside the shaded overlap volume
in $k$-space of the two spherical shells in Fig. 1 whose center-to-center
separation is $\mathbf{K\equiv k}_{1}+\mathbf{k}_{2}$, the total (or
center-of-mass) momentum wavevector for the pair.\ Otherwise, there is no
interaction and hence no pairing. For neutral-fermion systems there is no
cutoff $\hbar \omega _{D}$ and $V$ is \textit{not} constant in $\mathbf{k}$
and $\mathbf{k}^{\prime }$, except when the interaction range is much
shorter than the average fermion spacing as in a trapped Fermi gas where it
has become customary to employ contact (or $\delta $) interactions. Such are
a special case of the more general finite-ranged, separable interaction, as
is (\ref{BCSint}), introduced in Ref. \cite{Nozieres}, and for which the
conclusions below will continue to apply.

Vectors $\mathbf{k}$\ ending at all points of a simple-cubic lattice in $k$%
-space\ (in the 3D case) with lattice spacing $2\pi /L$ with $L$ the system
size, ensure that the interaction (\ref{BCSint})\ is nonzero provided such
points are in the shaded overlap volume of Figure 1. The endpoints of two
vectors $\mathbf{k}$\ and $\mathbf{k}^{\prime }$\ are illustrated in Figure
2. In the thermodynamic limit there are infinitely many acceptable $\mathbf{k%
}$\ values allowed for each fixed value of $\mathbf{K}$ for the $K\geq 0$ CP
eigenvalue equation \cite{Cooper}\cite{PC98} with interaction (\ref{BCSint}%
), namely 
\begin{equation}
V\sum_{\mathbf{k}}{}^{^{\prime }}(\hbar ^{2}k^{2}/m+\hbar
^{2}K^{2}/4m-2E_{F}-\mathcal{E}_{K})^{-1}=1.  \label{CPKeqn}
\end{equation}%
\begin{figure}[t]
\begin{center}
\includegraphics{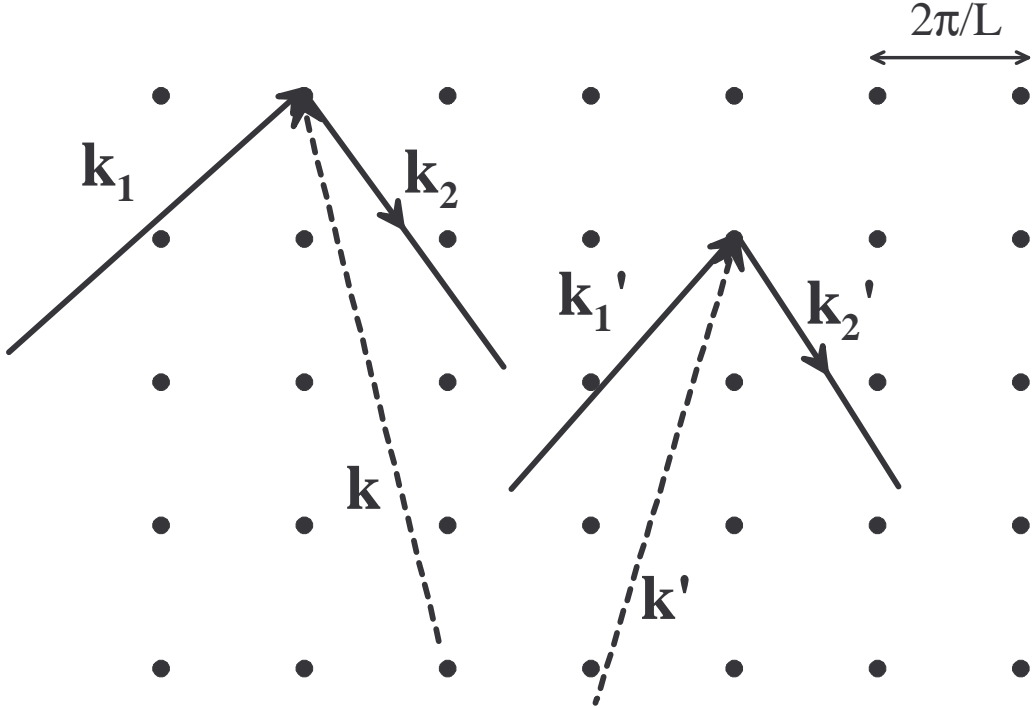}
\end{center}
\caption{Two distinct relative-momentum wavevector values $\mathbf{k}\equiv {%
\frac{1}{2}}(\mathbf{k}_{1}-\mathbf{k}_{2})$\ and $\mathbf{k}^{\prime
}\equiv \frac{1}{2}\mathbf{(k}_{1}^{\prime }-\mathbf{k}_{2}^{\prime }\mathbf{%
)}$ (dashed lines) whose tips are within the overlap volume of Figure 1.
They contribute to the summation in (\ref{CPKeqn}) or to the integral in (%
\ref{mCP}) to determine a CP energy $\mathcal{E}_{K}$. This illustrates why
the $\protect\delta _{\mathbf{k,k}^{\prime }}$ term for BCS pairs in (\ref%
{bb+}) is \textit{always }zero for CPs, with the Pauli Principle being
strictly satisfied.}
\label{BCSandCPsf2}
\end{figure}
The prime on the summation sign signifies the restrictions in (\ref{BCSint})
using $k_{F}$ as lower limit. For CPs consisting of equal and opposite
momentum electrons $K=0$, solving (\ref{CPKeqn})\ analytically gives the
familiar result of infinite-lifetime, negatively-bound (with respect to $%
2E_{F}$) CPs 
\begin{equation}
\mathcal{E}_{0}=-2\hbar \omega _{D}/[\exp \{2/N(E_{F})V\}-1]%
\mathrel{\mathop{\longrightarrow }\limits_{\lambda \rightarrow 0}}-2\hbar
\omega _{D}e^{-2/\lambda }  \label{CooperE}
\end{equation}%
where $\lambda \equiv N(E_{F})V\geq 0$ with $N(E_{F})$ the density of
electronic states at the Fermi energy. The first expression for $\mathcal{E}%
_{0}$ is exact for all values $\lambda $ in 2D [where $N(\epsilon )$ is
constant] and is otherwise a good approximation if $\hbar \omega _{D}\ll
E_{F}$ as occurs in metals.\ If hole CPs are included in the BS scheme along
with electron CPs as in (\ref{CPKeqn}) and (\ref{CooperE}), instead of (\ref%
{CooperE}) one obtains the evidently unphysical purely-imaginary result%
\begin{equation}
\mathcal{E}_{0}=\pm i2\hbar \omega _{D}/\sqrt{e^{2/\lambda }-1}
\label{pureimagCP}
\end{equation}%
(see detailed treatment in Ref. \cite{deLlRev}). Replacing the IFG sea by
the BCS ground-state sea, instead of (\ref{CPKeqn}) the eigenvalue equation
for $\mathcal{E}_{K}$ in 2D \cite{EPJD} (and a very similar one in 3D \cite%
{Honolulu} and in 1D \cite{1D}) is rather 
\begin{eqnarray}
\frac{1}{2\pi }\lambda \hbar
v_{F}\int_{k_{F}-k_{D}}^{k_{F}+k_{D}}dk\int_{0}^{2\pi }d\varphi u_{\mathbf{K}%
/2+\mathbf{k}}v_{\mathbf{K}/2-\mathbf{k}}\times \{u_{\mathbf{K}/2-\mathbf{k}%
}v_{\mathbf{K}/2+\mathbf{k}}-u_{\mathbf{K}/2+\mathbf{k}}v_{\mathbf{K}/2-%
\mathbf{k}}\}\times  &&  \nonumber \\
\times \frac{E_{\mathbf{K}/2+\mathbf{k}}+E_{\mathbf{K}/2-\mathbf{k}}}{-%
\mathcal{E}_{K}^{2}+(E_{\mathbf{K}/2+\mathbf{k}}+E_{\mathbf{K}/2-\mathbf{k}%
})^{2}} &=&1  \label{mCP}
\end{eqnarray}%
where $v_{F}\equiv \sqrt{2E_{F}/m}$, $\varphi $ is the angle between $%
\mathbf{K}$ and $\mathbf{k}$, $k_{D}$ is defined by $k_{D}/k_{F}\equiv \hbar
\omega _{D}/2E_{F}\ll 1$, $E_{\mathbf{k}}\equiv \sqrt{\xi _{k}{}^{2}+\Delta
^{2}}$ with $\xi _{k}\equiv \hbar ^{2}k^{2}/2m-E_{F}$ and$\ \Delta $ is the
electronic gap, while $v_{k}^{2}\equiv {\frac{1}{2}}(1-\xi _{k}/E_{\mathbf{k}%
})$ and $u_{k}^{2}\equiv 1-v_{k}^{2}$ are the usual Bogoliubov functions %
\cite{Bog}. Note that the lower limit in (\ref{BCSint}) is being taken as $%
\sqrt{k_{F}^{2}-k_{D}^{2}}$ as in BCS theory \cite{BCS} which does not
exclude hole pairings. Besides endowing CPs with the physically expected 
\textit{finite} lifetimes, the more general BS treatment yields the
nontrivial eigenvalue solution for $K=0$ given by $\mathcal{E}_{0}=2\Delta $%
, with $\Delta $ $\equiv $\ $\hbar \omega _{D}/\sinh (1/\lambda )%
\mathrel{\mathop{\longrightarrow }\limits_{\lambda
\rightarrow 0}}2\hbar \omega _{D}e^{-1/\lambda }$\ the single-electron BCS
energy gap, as opposed to the ordinary CP problem giving (\ref{CooperE}).
[Note the factor of two difference in the exponential compared with (\ref%
{CooperE}) due to excluding \cite{Cooper} or not \cite{BCS} hole pairs.] The
nontrivial (new) BS solution governed by (\ref{mCP})\ has been called \cite%
{Honolulu}-\cite{1D} a ``moving CP'' composite-boson excitation mode to
distinguish it from another (a trivial or known) sound wave solution
sometimes called the Anderson-Bogoliubov-Higgs (\cite{bts} p. 44), \cite{ABH}%
, \cite{Higgs} excitation mode which in contrast with the moving CP mode
vanishes as $K\rightarrow 0$ for any finite coupling, i.e., is gapless, and
is governed by another eigenvalue equation \cite{Honolulu}\cite{EPJD}\
different from (\ref{mCP}). In the pure boson gas, on the other hand, both
``particle'' and ``sound'' solutions are indistinguishable \cite{Gavoret}%
\cite{Hohenberg}. In the fermion case of interest here it should be feasible
to search for, identify and distinguish both particle and sound modes
experimentally, e.g., see Ref. \cite{CG}.

\section{BCS pairs with nonzero total momentum}

Whether the pairwise interfermion interaction is between charge carriers or
between neutral atoms, a CP state of energy $\mathcal{E}_{K}$,\ defined via
eigenvalue equation (\ref{CPKeqn}) or (\ref{mCP}),\ will clearly be
characterized \textit{only by a definite} $\mathbf{K}$ but \textit{not }%
definite $\mathbf{k}$. Although elementary, this is our main point. It
contrasts with that of a ``BCS pair'' defined as a dimer with fixed\ $%
\mathbf{K}$ and $\mathbf{k}$ (or equivalently fixed $\mathbf{k}_{1}$\ and $%
\mathbf{k}_{2}$), even though only the case $\mathbf{K}=0$ is considered in
Ref. \cite{BCS}, for which their annihilation $b_{\mathbf{k}}$ and creation $%
b_{\mathbf{k}}^{\dag }$ operators are not quite bosonic since they obey the
relations, Ref. \cite{BCS} Eqs. (2.11) to (2.13), 
\begin{equation}
\left[ b_{\mathbf{k}},b_{\mathbf{k}^{\prime }}^{\dag }\right] =(1-n_{-%
\mathbf{k}\downarrow }-n_{\mathbf{k}\uparrow })\delta _{\mathbf{kk}^{\prime
}}  \label{bb+0}
\end{equation}%
\begin{equation}
\left[ b_{\mathbf{k}}^{\dag },b_{\mathbf{k}^{\prime }}^{\dag }\right] =\left[
b_{\mathbf{k}},b_{\mathbf{k}^{\prime }}\right] =0  \label{bb0}
\end{equation}%
where $n_{{_{\mathbf{\pm k}s}}}\equiv {a_{\pm \mathbf{k}s}^{\dag }}{a_{\pm 
\mathbf{k}s}}$ are fermion number operators, with creation $a_{\mathbf{k_{1}}%
s}^{\dag }$\ and annihilation $a_{\mathbf{k_{1}}s}$\ operators referring to
the fermions, and 
\begin{equation}
\left\{ b_{\mathbf{k}},b_{\mathbf{k}^{\prime }}\right\} =2b_{\mathbf{k}}b_{%
\mathbf{k}^{\prime }}(1-\delta _{\mathbf{kk}^{\prime }})
\label{pseudofermion0}
\end{equation}%
which is not quite fermionic, unless $\mathbf{k}=\mathbf{k}^{\prime }$ when (%
\ref{bb+}) is not bosonic. The precise Bose commutation relations are, of
course,%
\begin{eqnarray}
\left[ b_{\mathbf{k}},b_{\mathbf{k}^{\prime }}^{\dag }\right] &=&\delta _{%
\mathbf{kk}^{\prime }}  \label{bosecomm} \\
\left[ b_{\mathbf{k}}^{\dag },b_{\mathbf{k}^{\prime }}^{\dag }\right] &=&%
\left[ b_{\mathbf{k}},b_{\mathbf{k}^{\prime }}\right] =0  \label{bosecomm2}
\end{eqnarray}%
with (\ref{bosecomm}) differing sharply from (\ref{bb+0}). The fermion
creation $a_{\mathbf{k_{1}}s}^{\dag }$\ and annihilation $a_{\mathbf{k_{1}}%
s} $\ operators satisfy the usual Fermi anti-commutation relations%
\begin{eqnarray}
\left\{ a_{\mathbf{k_{1}}s}^{\dag },a_{\mathbf{k_{1}^{\prime }}s^{\prime
}}^{\dagger }\right\} &=&\left\{ a_{\mathbf{k_{1}}s},a_{\mathbf{%
k_{1}^{\prime }}s^{\prime }}\right\} =0  \label{3} \\
\left\{ a_{\mathbf{k_{1}}s},a_{\mathbf{k_{1}^{\prime }}s^{\prime }}^{\dag
}\right\} &=&\delta _{\mathbf{k_{1}k_{1}^{\prime }}}\delta _{ss^{\prime }}.
\label{3a}
\end{eqnarray}

The distinction between BCS pairs and CPs holds for the original or
``ordinary'' CPs \cite{Cooper} as in (\ref{CPKeqn}) and (\ref{CooperE}). It
also applies to the generalized BS CPs defined\ more consistently without
excluding two-hole pairs when the lower limit in (\ref{BCSint}) is taken as $%
\sqrt{k_{F}^{2}-k_{D}^{2}}$ as in BCS theory \cite{BCS}.

The BCS-pair annihilation and creation operators for any $\mathbf{K>}$ $0$
can be defined quite generally as 
\begin{equation}
b_{\mathbf{kK}}\equiv a_{\mathbf{k}_{2}\downarrow }a_{\mathbf{k}_{1}\uparrow
}\hbox{
\ \ \ \ and \ \ \ \ }b_{\mathbf{kK}}^{\dag }\equiv a_{\mathbf{k}_{1}\uparrow
}^{\dag }a_{\mathbf{k}_{2}\downarrow }^{\dag }  \label{4}
\end{equation}%
where $a_{\mathbf{k_{1}}s}^{\dag }$\ and $a_{\mathbf{k_{1}}s}$ obey (\ref{3}%
) and (\ref{3a}), and\ as before $\mathbf{k}\equiv {{\frac{1}{2}}}(\mathbf{%
k_{1}-k_{2})}$ and $\mathbf{K}\equiv \mathbf{k_{1}}+\mathbf{k_{2}}$ are the
relative and total (or center-of-mass) momentum wavevectors, respectively,
associated with two fermions with wavevectors 
\begin{equation}
\mathbf{k}_{1}=\mathbf{K}/2+\mathbf{k}\ \hbox{\ \ \ \ \ and \ \ \ \ }\mathbf{%
\ k}_{2}=\mathbf{K}/2-\mathbf{k}.  \label{6}
\end{equation}%
Using the same techniques to derive (\ref{bb+0}) to (\ref{pseudofermion0})
valid for $\mathbf{K}=0$, the operators $b_{\mathbf{kK}}$ and $b_{\mathbf{kK}%
}^{\dag }$ are found\ to satisfy: a) the ``pseudo-boson'' commutation
relations 
\begin{equation}
\left[ b_{\mathbf{kK}},b_{\mathbf{k}^{\prime }\mathbf{K}}^{\dag }\right]
=(1-n_{\mathbf{K}/2-\mathbf{k}\downarrow }-n_{\mathbf{K}/2+\mathbf{k}%
\uparrow })\delta _{\mathbf{kk}^{\prime }}  \label{bb+}
\end{equation}%
\begin{equation}
\left[ b_{\mathbf{kK}}^{\dag },b_{\mathbf{k}^{\prime }\mathbf{K}}^{\dag }%
\right] =\left[ b_{\mathbf{kK}},b_{\mathbf{k}^{\prime }\mathbf{K}}\right] =0
\label{bb}
\end{equation}%
where $n_{{_{\mathbf{K/2\pm k}s}}}\equiv {a_{\mathbf{K}/2\pm \mathbf{k}%
s}^{\dag }}{a_{\mathbf{K}/2\pm \mathbf{k}s}}$ are fermion number operators;
as well as b) a ``pseudo-fermion'' anti-commutation relation 
\begin{equation}
\left\{ b_{\mathbf{kK}},b_{\mathbf{k}^{\prime }\mathbf{K}}\right\} =2b_{%
\mathbf{kK}}b_{\mathbf{k}^{\prime }\mathbf{K}}(1-\delta _{\mathbf{kk}%
^{\prime }}).  \label{pseudofermion}
\end{equation}%
Our only restriction was that $\mathbf{K\equiv k}_{1}+\mathbf{k}_{2}=\mathbf{%
k}_{1}^{\prime }+\mathbf{k}_{2}^{\prime }$. If $\mathbf{K}=0$ so that $%
\mathbf{k}_{1}=-\mathbf{k}_{2}=\mathbf{k}$ (the only case considered by
BCS), and calling $b_{\mathbf{kK=0}}\equiv b_{\mathbf{k}}$, etc., (\ref{bb+}%
) to (\ref{pseudofermion}) become (\ref{bb+0}) to (\ref{pseudofermion0}), as
they should. So, neither BCS pairs with $\mathbf{K\geq }$ $0$\ are\textit{\ }%
bosons as the relation (\ref{bb+}) contains additional terms not \cite%
{Schrieffer64}\ present in the the usual boson commutation relations
analogous to (\ref{bosecomm}).

To our knowledge, no one has yet succeeded in constructing CP creation and
annihilation operators that obey Bose commutation relations, starting from
fermion creation $a_{\mathbf{k_{1}}s}^{\dag }$\ and annihilation $a_{\mathbf{%
k_{1}}s}$\ operators, as is \textit{postulated }in Refs. \cite{Tolma,PLA2}
in a generalized BEC theory that contains BCS theory as a special case. This
postulation is grounded in magnetic-flux quantization experiments\ \cite%
{classical}-\cite{cuprates} that establish the presence of charged
pairs---albeit without being able \cite{Gough}\ to specify the \textit{sign}
of those charges.\ However, although the eigenvalues of $b_{\mathbf{kK}%
}^{\dag }b_{\mathbf{kK}}$ are $0$ or $1$ in keeping with the Pauli Exclusion
Principle, those of ${\sum_{\mathbf{k}}}b_{\mathbf{kK}}^{\dag }b_{\mathbf{kK}%
}$ are evidently $0,1,2,...$ because of the indefinitely many values taken
on by the summation index $\mathbf{k.}$ This implies BE statistics and
corroborates the qualitative conclusions reached above. A discussion in
greater detail of this is found in Refs. \cite{FujitaMorabito}\cite{FG}.

\section{Conclusion}

In conclusion, in the thermodynamic limit \textit{any} number of CPs with a
definite total (or center-of-mass) wavevector $\mathbf{K}$ can occupy a 
\textit{single} state of CP energy $\mathcal{E}_{K}$ and thus obey BE
statistics. Hence, CPs can undergo a BEC. All this holds for \textit{any}
coupling, i.e., regardless of the size of the CPs and of their mutual
overlap. As in liquid $^{4}$He,\ the BEC statistical mechanism may serve as
a starting point---even before the effects of interboson interactions
(excluded also in BCS theory) are accounted \cite{Ceperley}\ for---in
constructing a microscopic theory of superconductivity, or of
neutral-fermion superfluidity, that will eventually lead to $T_{c}$ values
calculated from first principles\ without adjustable parameters\ and that
hopefully\ agree with observed ones.

{\large \textbf{Acknowledgments}}

MdeLl especially thanks John R. Clem for encouraging us to write this paper,
as well as M. Casas and M. Moreno for discussions, E. Braun and S. Ram\'{\i}%
rez for invaluable input, and UNAM-DGAPA-PAPIIT (Mexico) for grant IN106401
as well as CONACyT (Mexico) for grant 41302 in partial support. He also
thanks the Texas Center for Superconductivity, University of Houston, for
its hospitality and support. MdeLl and FJS thank the NSF (USA) for partial
support through grant INT-0336343 made to the Consortium of the Americas for
Interdisciplinary Science, University of New Mexico, Albuquerque, NM, USA.


\begin{thebibliography}{99}
\bibitem{Samuelsson} P. Samuelsson and M. B\"{u}ttiker, Phys. Rev. Lett. 
\textbf{89} 046601 (2002).

\bibitem{Tolma} V.V. Tolmachev, Phys. Lett. A \textbf{266} 400 (2000).

\bibitem{PLA2} M. de Llano and V.V. Tolmachev, Physica A \textbf{317} 546
(2003).

\bibitem{Friedel} J. Labb\'{e}, S. Barisic, and J. Friedel, Phys. Rev. Lett. 
\textbf{19} 1039\ (1967).

\bibitem{Eagles} D.M. Eagles, Phys. Rev. \textbf{186} 456 (1969).

\bibitem{Leggett} A.J. Leggett, in \textit{Modern Trends in the Theory of
Condensed Matter, }ed. by A. Pekalski and J. Przystawa, Lecture\ Notes in
Physics, Vol. 115, (Springer-Verlag, Berlin, 1980); J. Phys. (Paris) Colloq. 
\textbf{41} C7-19\ (1980).

\bibitem{Miyake} K. Miyake, Prog. Theor. Phys. \textbf{69} 1794 (1983).

\bibitem{Nozieres} P. Nozi\`{e}res and S. Schmitt-Rink, J. Low. Temp. Phys. 
\textbf{59} 195 (1985).

\bibitem{Randeria89} M. Randeria, J.-M. Duan, and L.-Y. Shieh, Phys. Rev.
Lett. \textbf{62} 981\ (1989); \textbf{62} 2887(E) (1989); Phys. Rev. B 
\textbf{41}, 327\ (1990).

\bibitem{vanderMarel} D. van der Marel, Physica C \textbf{165} 35 (1990).

\bibitem{DZ92} M. Drechsler and W. Zwerger, Ann. der Physik \textbf{1} 15
(1992).

\bibitem{Haus} R. Haussmann, Z. Phys. B \textbf{91} 291 (1993); Phys. Rev. B 
\textbf{49} 12975 (1994).

\bibitem{Pistolesi} F. Pistolesi and G.C. Strinati, Phys. Rev. B \textbf{49}
6356 (1994); Phys. Rev. B \textbf{53} 15168 (1996).

\bibitem{PRB94} M. Casas, J.M. Getino, M. de Llano, A. Puente, R.M. Quick,
H. Rubio \& D.M. van der Walt, Phys. Rev. B \textbf{50} 15945 (1994).

\bibitem{PRB95} R.M. Carter, M. Casas, J.M. Getino, M. de Llano, A. Puente,
H. Rubio, and D.M. van der Walt, Phys. Rev. B \textbf{52} 16149 (1995).

\bibitem{Cooper} L.N. Cooper, Phys. Rev. \textbf{104} 1189 (1956).

\bibitem{BCS} J. Bardeen, L.N. Cooper, and J. R Schrieffer, Phys. Rev. 
\textbf{108} (1957) 1175.

\bibitem{PT} J. Bardeen, Phys. Today (Jan. 1963) p. 25.

\bibitem{Feymann} R.P. Feynman, \textit{The Feynman Lectures on Physics},
(Addison-Wesley, Reading, 1965) pp. 7, 18, 21.

\bibitem{Josephson} B.D. Josephson, Rev. Mod. Phys. \textbf{36}, 216 (1964).

\bibitem{CMT02} J. Batle, M. Casas, M. Fortes, M. de Llano, F.J. Sevilla,
M.A. Sol\'{\i}s, and V.V. Tolmachev, Cond. Matter Theories \textbf{18} 111
(2003). Or cond-mat/0211456.

\bibitem{deLlRev} M. de Llano, in \textit{Frontiers in Superconductivity
Research}, ed. by B.P. Martins (Nova Science Publishers, NY, 2004) \ Also in
cond-mat/0405071.

\bibitem{He3} D. Vollhardt and P. W\"{o}lfle, \textit{The Superfluid Phases
of Helium 3} (Taylor \& Francis, London, 1990).

\bibitem{Holland} M.J. Holland, B. DeMarco, and D.S. Jin, Phys. Rev. A 
\textbf{61} 053610 (2000) and refs. therein.

\bibitem{Li6} K.M. O'Hara, S.L. Hemmer, M.E. Gehm, S.R. Granade, and J.E.
Thomas, Science \textbf{298} 2179 (2002).

\bibitem{Holland01} M.J. Holland, S.J.J.M.F. Kokkelmans, M.L. Chiofalo, and
R. Walser, Phys. Rev. Lett. \textbf{87} 120406 (2001).

\bibitem{EddyT01} E. Timmermans, K. Furuya, P.W. Milonni, and A.K. Kerman,
Phys. Lett. A \textbf{285} 228 (2001).

\bibitem{Chiofalo02} M.L. Chiofalo, S.J.J.M.F. Kokkelmans, J.N. Milstein,
and M.J. Holland,\ Phys. Rev. Lett. \textbf{88} 090402 (2002).

\bibitem{Griffin02} Y. Ohashi and A. Griffin, Phys. Rev. Lett. \textbf{89}
130402 (2002).

\bibitem{Stringari02} L. Pitaevskii and S. Stringari, Science \textbf{298}
2144\textbf{\ }(2002).

\bibitem{Jin} C.A. Regal, C. Ticknor, J.L. Bohn, and D.S. Jin, Nature 
\textbf{424} 47 (2003).

\bibitem{Hulet} K.E. Strecker, G.B. Partridge, and R.G. Hulet, Phys. Rev.
Lett. \textbf{91}, 080406 (2003).

\bibitem{Jochim} S. Jochim, M. Bartenstein, A. Altmeyer, G. Hendl, C. Chin,
J. Hecker Denschlag, and R. Grimm, Phys. Rev. Lett. \textbf{91}, 240402
(2003).

\bibitem{Ketterle} M.W. Zwierlein, C. A. Stan, C. H. Schunck, S.M. F.
Raupach, S. Gupta, Z. Hadzibabic, and W. Ketterle, Phys. Rev. Lett. \textbf{%
91}, 250401 (2003).

\bibitem{bts} N.N. Bogoliubov, V.V. Tolmachev, and D.V. Shirkov, Fortschr.
Phys. \textbf{6} 605 (1958); and \textit{A New Method in the Theory of
Superconductivity} (Consultants Bureau, NY, 1959).

\bibitem{AGD} A.A. Abrikosov, L.P. Gorkov, and I.E. Dzyaloshinskii, \textit{%
Methods of Quantum Field in Statistical Physics }(Dover, NY, 1975) \S\ 33.

\bibitem{Honolulu} M. Fortes, M.A. Sol\'{\i}s, M. de Llano, and V.V.
Tolmachev, Physica C \textbf{364} 95 (2001).

\bibitem{EPJD} V.C. Aguilera-Navarro, M. Fortes, and M. de Llano, Sol. St.
Comm. \textbf{129} 577 (2004).

\bibitem{1D} M. Fortes, M. de Llano, and M.A. Sol\'{\i}s, to be published.

\bibitem{Lanzara} A. Lanzara, P.V. Bogdanov, X.J. Zhou, S.A. Kellar, D.L.
Feng, E.D. Lu, T. Yoshida, H. Eisaki, A. Fujimori, K. Kishio, J.-I.
Shimoyama, T. Noda, S. Uchida, Z. Hussain, and Z.-X. Shen, Nature \textbf{412%
} 510 (2001).

\bibitem{Cuk} T. Cuk\textit{, }F. Baumberger, D.H. Lu, N. Ingle, X.J. Zhou,
H. Eisaki, N. Kaneko, Z. Hussain, T.P. Devereaux, N. Nagaosa, and Z.X. Shen,%
\textit{\ }Phys. Rev. Lett. \textbf{93} 117003 (2004). Also in
cond-mat/0403521.

\bibitem{Hwang} J. Hwang, T. Timusk, and G.D. Gu, Nature \textbf{427} 714
(2004).

\bibitem{Kulic} M.L. Kulic, AIP Conf. Proc. \textbf{715}, 75 2004). Also in
cond-mat/0404287.


\bibitem{PC98} M. Casas, S. Fujita, M. de Llano, A. Puente, A. Rigo, and
M.A. Sol\'{\i}s, Physica C \textbf{295} 93 (1998).

\bibitem{Bog} N.N. Bogoliubov, N. Cim.\textbf{\ 7} 794 (1958).

\bibitem{ABH} {P.W. Anderson, Phys. Rev. }\textbf{112} 1900{\ (1958).}

\bibitem{Higgs} {P.W. Higgs, Phys. Lett. }\textbf{12} 132 {(1964).}

\bibitem{Gavoret} J. Gavoret and P. Nozi\`{e}res, Ann. Phys. (NY) \textbf{28}
349 (1964).

\bibitem{Hohenberg} P.C. Hohenberg and P.C. Martin, Ann. Phys. (NY) \textbf{%
34} 291 (1965).

\bibitem{CG} R.V. Carlson and A.M. Goldman, Phys. Rev. Lett. \textbf{31} 880
(1973) and\textbf{\ 34} 11 (1975).{\ }

\bibitem{Schrieffer64} J.R. Schrieffer, \textit{Theory of Superconductivity}
(Benjamin, New York, 1964) p. 38.

\bibitem{classical} B.S. Deaver, Jr. and W.M. Fairbank, Phys. Rev. Lett. 
\textbf{7}, 43 (1961).

\bibitem{classical2} R. Doll and M. N\"{a}bauer, Phys. Rev. Lett. \textbf{7}%
, 51 (1961).

\bibitem{cuprates} C.E. Gough, M.S Colclough, E.M. Forgan, R.G. Jordan, M.
Keene, C.M. Muirhead, I.M. Rae, N. Thomas, J.S. Abell, and S. Sutton, Nature%
\textbf{\ 326, }855 (1987).

\bibitem{Gough} C.E. Gough, priv. comm.

\bibitem{FujitaMorabito} S. Fujita and D.L. Morabito, Mod. Phys. Lett. B 
\textbf{12}, 753 (1998).

\bibitem{FG} S. Fujita and S. Godoy, \textit{Theory of High Temperature
Superconductivity} (Kluwer, NY, 2001) pp. 97-98.

\bibitem{Ceperley} P. Gr\"{u}ter, D. Ceperley, and F. Lalo\"{e}, Phys. Rev.
Lett. \textbf{79} 3549 (1997).
\end{thebibliography}
\end{document}